\documentclass{MIPRO}

\usepackage{cite}
\usepackage{amsmath,amssymb,amsfonts}
\usepackage{algorithmic}
\usepackage{graphicx}
\usepackage{textcomp}
\usepackage{xcolor}
\usepackage[T1]{fontenc}
\usepackage{flushend}
\usepackage{multirow}
\usepackage{multicol}
\usepackage{array}
\usepackage{url}
\usepackage[inline]{enumitem}
\usepackage[numbers]{natbib}
\usepackage{pgf-pie}
\usepackage{listings}
\usepackage{hyperref}

\newcommand\copyrighttext{%
  \footnotesize \textcopyright 2022 IEEE. Personal use of this material is permitted. Permission from IEEE must be obtained for all other uses, in any current or future media, including reprinting/republishing this material for advertising or promotional   purposes, creating new collective works, for resale or redistribution to servers or lists, or reuse of any copyrighted component of this work in other works. DOI: \href{https://doi.org/10.23919/MIPRO55190.2022.9803570}{10.23919/MIPRO55190.2022.9803570}}
\newcommand\copyrightnotice{%
\begin{tikzpicture}[remember picture,overlay]
\node[anchor=south,yshift=10pt] at (current page.south) {\fbox{\parbox{\dimexpr\textwidth-\fboxsep-\fboxrule\relax}{\copyrighttext}}};
\end{tikzpicture}%
}

\begin{document}

% Limit bibliography to max. 3 authors before trimming to et. al.
\bstctlcite{IEEEexample:BSTcontrol}

\title{Systematic review of automatic translation of high-level security policy into firewall rules}

\author{
    \IEEEauthorblockN{
    Ivan Kovačević\IEEEauthorrefmark{1},
    Bruno Štengl\IEEEauthorrefmark{2},
    Stjepan Groš\IEEEauthorrefmark{3}
    }

    \IEEEauthorblockA{
    University of Zagreb Faculty of Electrical Engineering and Computing, Zagreb, Croatia
    }

    \IEEEauthorrefmark{1}ivan.kovacevic@fer.hr,
    \IEEEauthorrefmark{2}bruno.stengl@fer.hr,
    \IEEEauthorrefmark{3}stjepan.gros@fer.hr
}

\maketitle

\copyrightnotice

\begin{abstract}
Firewalls are security devices that perform network traffic filtering. They are ubiquitous in the industry and are a common method used to enforce organizational security policy. Security policy is specified on a high level of abstraction, with statements such as “web browsing is allowed only on workstations inside the office network”, and needs to be translated into low-level firewall rules to be enforceable. There has been a lot of work regarding optimization, analysis and platform independence of firewall rules, but an area that has seen much less success is automatic translation of high-level security policies into firewall rules. In addition to improving rules’ readability, such translation would make it easier to detect errors.
% and has the potential to enhance cyber security education by .

This paper surveys of over twenty papers that aim to generate firewall rules according to a security policy specified on a higher level of abstraction. It also presents an overview of similar features in modern firewall systems. Most approaches define specialized domain languages that get compiled into firewall rule sets, with some of them relying on formal specification, ontology, or graphical models. The approaches' have improved over time, but there are still many drawbacks that need to be solved before wider application.
\end{abstract}

\renewcommand\IEEEkeywordsname{Keywords}
\begin{IEEEkeywords}
\textit{network security, security policy, firewall}
\end{IEEEkeywords}

\section{INTRODUCTION}

Security policy defines a system's security requirements trough standards, rules, and practices \cite{schuba1997reference}. In firewall implementations, security policies are implemented using a set of firewall rules that match network packets and define actions that are performed over such packets. In traditional firewalls, rules are written in domain specific languages that use various low-level technical details, such as IP addresses, ports, and protocols, which makes them challenging to define and manage. \citeauthor{pop00121}~\cite{pop00121} compares the readability of such rules to assembly code.

Organizations would benefit from the ability to describe security policies on a higher level of abstraction, where they would be more concerned with the semantics of what they want to achieve rather than low-level technical details. For instance, a single high-level policy statement could allow all workstations access to domain controller (DC) services, such as offered by Active Directory, instead of having multiple rules that filter packets according to source subnets, server IP address and characteristic DC service ports. Such policy definitions would be more comprehensive and much easier to understand and manage, especially in large Enterprise networks.
%TODO: motivacija, ciljevi (Generator ITS-a \citeauthor{9695437}~\cite{9695437}, biljeznica)
Furthermore, it would be useful to be able to enforce such policies using existing firewall systems that companies already use.
Another area that would benefit from such capabilities are approaches that generate models of IT systems for cyber security exercises, such as \cite{9695437}, where abstract policies need to be implemented on concrete firewall systems used during a cyber exercise.

% A number of approaches appeared that aim to solve this problems. They define security policies on a higher level of abstraction, and according to those definitions automatically build a set of appropriate low-level firewall rules. In parallel, a new generation of firewall systems introduced various high-level concepts that enable a more direct mapping of policy concepts into rules.

% Olakšavanje implementacije sigurnosnih politika u velikim mrežama koje koriste postojeća firewall rješenja
% Konstatirati definiciju politike visoke razine (članci su ju već definirali) i pravila, konfiguracija firewalla
% Smanjenje mogućnosti pogreške
% Implementacija politika dobivenih iz sustava poput Generatora ITS-a

In accordance with the aforementioned motivation, our survey aims to find approaches that generate firewall configurations based on security policies defined at a high level of abstraction. Here, \textit{high level of abstraction} refers to the fact that definitions of such policies avoid relying on network-related technical details, including various aliases thereof (e.g \textit{home\_network}). Instead, they use formulations closer to natural language, or an abstract visual definition trough a graphical user interface (GUI). In addition, the approaches' must provide a concrete implementation, and not be limited to a proposal or patent only describing a framework or idea.
Other types of approaches, such as those
%using definitions of high-level security policies for
focusing on firewall rule verification and error detection, are outside the scope of this paper. Last but not least, our survey focuses only on papers for which the full text is available in English.

In parallel to the aforementioned approaches, next generation firewall (NGFW) systems~\cite{pescatore2009defining} introduce various high-level concepts that enable a more direct mapping between policies and firewall rules. This survey includes a brief overview of their features, focusing on those which mirror functionality of the surveyed approaches.

This paper is organized as follows. Section \ref{sec-method} describes the methods used to perform the survey. Next, Section \ref{sec-results} provides an overview of the surveyed papers, and shows examples of similar functionality offered by modern firewall systems. The results and their significance are discussed in Section \ref{sec-discussion}.
Finally, the paper presents related work in Section \ref{sec-related-work} and wraps up with conclusions in Section \ref{sec-conclusion}.

\section{Methods}
\label{sec-method}

This section describes the methods used to survey the literature. Several literature searches were performed using Google Scholar and Semantic Scholar, supported by the software tool Publish or Perish~\cite{HarzingPoP}. The following search terms were used:
\begin{enumerate*}[label=(\roman*)]
    \item \textit{high-level firewall rules},
    \item \textit{high-level firewall policies},
    \item \textit{management of firewall rules},
    \item \textit{management of firewall policies}, and
    \item \textit{generating firewall rules from high-level policies}
\end{enumerate*}.
From each search, $50$ most relevant results were collected, yielding $500$ papers in total. After removing duplicate papers, $304$ papers remained. Out of these, only $23$ papers representing $17$ distinct approaches remained within the scope of this survey.
%(* brojke će biti osvježene poslije popunjavanja tablice, kada se još neki izbace iz scopea).
The remaining approaches were analyzed according to research questions listed below. Each question is designated with a label and describes potential answers where necessary:

\begin{enumerate}
    \item \textit{Level (H/M/L)}: What is the level of abstraction of security policies defined as inputs to the approach? Possible values include the following:
    \begin{itemize}
        \item H: Policies are defined using natural language, as spoken by management and policy makers.
        \item M: Policies are defined using formal specification or graphs with concepts like users, types of applications, types of roles, etc. They do not primarily focus on technical details such as IP addresses, ports, etc., and instead link such details using a knowledge base and/or separate configuration.
        \item L: Defines policies with technical details that are hidden behind concepts similar to roles, instead of being stated as low-level firewall rules. An example would be defining policies related to the \textit{public\_web\_server} role, and separately label several servers with the aforementioned role.
    \end{itemize}
    \item \textit{KB (yes/no/partial/N.A.)}: Does the approach include a predefined knowledge base, or predefined rule conversion (expert) rules? If large parts of the knowledge base must be tailored specifically for the organizational IT system where they are to be deployed, it is considered as a partial KB.
    \item \textit{GUI (yes/no/N.A.)}: Does the approach include a GUI and/or a visualization component?
    \item \textit{Output rules}: Which types of low-level firewall rules can the approach generate? Examples could include iptables rules and Cisco firewall rules.
    \item \textit{Category}: What is the central concept used to define and process policies? Possible categories include the following:
    \begin{itemize}
        \item \textit{Ontology}: the approach uses ontologies to define the high-level security policy.
        %\item \textit{Expert system}: the approach provides a knowledge base with some type of expert rules or scripts that generate firewall rules (distinct to ontology).
        \item \textit{Language}: the approach is based on features characteristic to high-level programming languages, such as operators or inheritance, or defines policies using markup languages such as XML.
        \item \textit{Formal}: the approach uses formal predicates to define the security policy, e.g. as a collection of conditions that must always be met. Such formal specification of high-level policy is used to generate low-level rules using various automated reasoning and optimization algorithms.
        \item \textit{Graph}: the approach uses graphs, such as Petri nets, to link entities and define high-level policies.
        %\item \textit{Matrix model}: the approach uses matrices to link entities.
    \end{itemize}
    \item \textit{Usability}: Do authors report a usability study? If yes, what type? Possible types include the following:
    \begin{itemize}
        \item \textit{Study}: Authors report a usability study with target users, in which it was confirmed that the approach makes policies easier to manage than low-level rules and that its expressiveness is sufficient for real-world usage scenarios.
        \item \textit{Claims}: Authors describe experiments in which they define policies and claim that they were successful in using the system.
        \item \textit{N.A.}: Usability was not evaluated.
    \end{itemize}
\end{enumerate}

\section{RESULTS}
\label{sec-results}

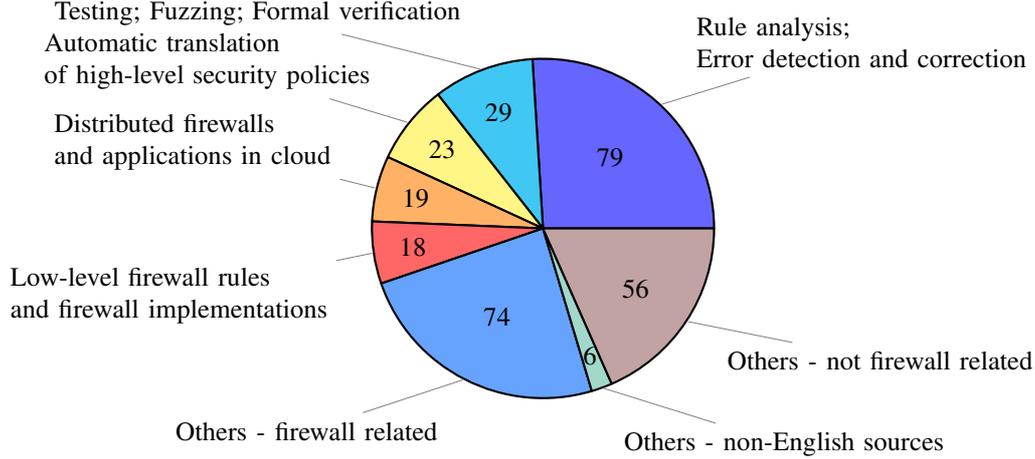
\begin{figure*}
  \centering
    \begin{tikzpicture}[scale=0.75]
    \pie[sum=auto, /tikz/every pin/.style={align=left}, text=pin]{
        79/Rule analysis;\\Error detection and correction, % 79/304
        29/Testing; Fuzzing; Formal verification, % 29/304
        23/Automatic translation\\of high-level security policies, % 23/304
        19/Distributed firewalls\\and applications in cloud,  % 19/304
        18/Low-level firewall rules\\and firewall implementations,  % 18/304
        74/Others - firewall related,  % 74/304
        6/Others - non-English sources,  % 6/304
        56/Others - not firewall related % 56/304
    }
    \end{tikzpicture}
    \caption{Rough categorization of initial search results. Duplicate results are excluded. Individual categories of papers are explained in Section \ref{sec-results}.}
    \label{fig-initial-chart}
\end{figure*}

By applying the method outlined in Section \ref{sec-method}, we initially found $304$ unique papers. We propose a rough informal categorization of those papers according to research areas as shown in Fig. \ref{fig-initial-chart}. Papers within the scope of this survey are labeled as \textit{Automatic translation of high-level security policies}, and their overview is provided in Section \ref{sec-results-in-scope}. Remaining categories are briefly explained in Section \ref{sec-results-outside}. There are many border cases in which papers could potentially be categorized into alternative categories, so it is possible that other researchers could obtain slightly different paper counts. Finally, Section \ref{sec-results-industry} provides an overview of comparable features of some modern firewall systems commonly used in the industry.

\subsection{Automatic translation of high-level security policies}
\label{sec-results-in-scope}

Some properties of the surveyed approaches are outlined in Table \ref{table-results}. Among these approaches, high-level security policies are frequently defined using a programming language inspired syntax or XML documents. Formal methods and graphical models are used to a lesser degree, with only one approach relying on an ontology. The following paragraphs provide brief examples of policy definitions for each category.

\begin{table*}[h]
\caption{Overview of the surveyed approaches. In cases in which approaches were published trough multiple papers, the year of the most recent one is indicated. Individual columns and categories are explained in Section \ref{sec-method}.}
\label{table-results}
\centering
\def\arraystretch{1.25}
\begin{tabular}{|l|c|ccccc|p{16em}|}
 \hline
 Approach & Year & Level & KB & GUI & Category & Usability & Output rules
 \\
  \hline \citeauthor{pop00355}~\cite{pop00355,pop00038} & 2004 & L & partial & yes & Language & Study & Lucent VPN, Check Point FireWall-1, Cisco PIX Firewall, Cisco IOS \\
 \hline \citeauthor{pop00083}~\cite{pop00083} & 2004 & M & no & no & Formal & N.A. & iptables, tested with netfilter \\
 \hline \citeauthor{pop00005}~\cite{pop00005} & 2007 & L & no & no & Language & Study & format similar to iptables \\
 \hline \citeauthor{pop00073}~\cite{pop00073} & 2007 & M & partial & yes & Graph & Study & Colored Petri net formalism \\
 \hline \citeauthor{pop00029}~\cite{pop00010,pop00029} & 2009 & L & no & no & Formal & Claims & N.A. \\
 \hline \citeauthor{pop00158}~\cite{pop00158} & 2009 & M & partial & no & Language & N.A. & N.A. (vendor specific) \\
 \hline \citeauthor{pop00087}~\cite{pop00087} & 2010 & H & yes & yes & Ontology & Claims & N.A. \\
 \hline \citeauthor{pop00071}~\cite{pop00241,pop00071} & 2011 & L & no & no & Language & Claims & iptables \\
 \hline \citeauthor{pop00153}~\cite{pop00153} & 2013 & L & no & no & Language & Study & Palo Alto PanOS \\
 \hline \citeauthor{pop00163}~\cite{pop00163,pop00062} & 2014 & L & partial & yes & Language & Study & format similar to iptables \\
 \hline \citeauthor{pop00021}~\cite{pop00021} & 2015 & L & no & no & Language & Claims & ClickOS-based firewall implementation \\
 \hline \citeauthor{pop00485}~\cite{pop00485} & 2015 & H & yes & yes & Language & N.A. & packet filtering virtual network function \\
 \hline \citeauthor{pop00059}~\cite{pop00059,pop00076} & 2016 & L & no & yes & Graph & Study & Packet filter \\
 \hline \citeauthor{pop00436}~\cite{pop00436} & 2016 & L & no & no & Formal & Claims & Netfilter \\
 \hline \citeauthor{pop00472}~\cite{pop00472} & 2019 & M & partial & no & Language & Study & iptables \\
 \hline \citeauthor{pop00013}~\cite{pop00013} & 2020 & M & no & no & Graph & N.A. & N.A. \\
 \hline \citeauthor{pop00065}~\cite{pop00065,pop00493} & 2020 & L & yes & no & Formal & Claims & N.A. \\
 \hline
\end{tabular}
\end{table*}

\citeauthor{pop00485}~\cite{pop00485} use sentences close to natural language to describe high-level security requirements. Subjects, objects and actions that can be used in sentences are pre-configured in the approaches' knowledge base. A few examples of such requirements are shown in Listing \ref{lst-pop00485}. The first one bans users from visiting gambling sites, and the second one allows user Alice to access Internet between 18:30 and 20:00 hours. The security requirements are automatically translated to configurations of appropriate network devices.

\begin{figure}[h]
    \centering
    \begin{lstlisting}[
        frame=single, basicstyle=\ttfamily,
        caption=Examples of high-level security requirements from \citeauthor{pop00485}~\cite{pop00485}.,
        label=lst-pop00485
    ]
    "do not access gambling sites"
    "allow Internet traffic
        from 18:30 to 20:00 for Alice"
    \end{lstlisting}
\end{figure}

\citeauthor{pop00153}~\cite{pop00153} defines a syntax similar to programming languages, which supports variables and inheritance, and can be used to define rules more comprehensive than traditional low-level firewall rules. Listing \ref{lst-pop00153} shows a simplified example where this syntax is used to describe the Facebook web application and define a rule that grants access to it from the internal network. In a similar manner, other applications and locations on the network can be described, which would enable specifying policy based on applications and groups of resources rather than IP addresses, ports and protocols.

\begin{figure}[h]
    \centering
    \begin{lstlisting}[
        frame=single, basicstyle=\ttfamily,
        caption=An example of a configured application and defined rule from \citeauthor{pop00153}~\cite{pop00153}.,
        label=lst-pop00153
    ]
    application facebook {
        protocol tcp
        port 80, 443
        signature "www.facebook.com"
    }
    rule "facebook access" {
        application facebook
        from internal
        to external
        source client -network
        destination any
        action allow
    }
    \end{lstlisting}
\end{figure}

\citeauthor{pop00436}~\cite{pop00436} define the security policy using high-level goals that need to be enforced. These goals are specified using formal predicates that describe desirable states of packets at certain locations in the network. The formally specified goals transparently support network address translation (NAT) and are used to automatically generate, optimize, and localize low-level firewall rules. Localization here refers to the task through which rules need to be placed on multiple firewalls throughout the network, with each having its own surroundings and potentially a different NAT configuration. Unfortunately, goals are stated using complex logical formulas and their notation does not seem to be very comprehensible.

\citeauthor{pop00059}~\cite{pop00059,pop00076} develop an educational tool, SP2Model, that can be used to model firewall rules in a graphical manner. After network resources have been defined, users are presented with a graphical interface where they can draw a graph that describes the security policy. This graph still retains low-level rule semantics, and it is questionable whether it could be applicable in larger networks.

\citeauthor{pop00013}~\cite{pop00013} propose a specialized solution for Kubernetes~\cite{kubernetes} clusters that automatically generates firewall rules according to defined services and their communication requirements. Users can manually allow additional communication flows. We classify this as a graph approach because the service definitions effectively form a graph, with services as nodes and their communication requirements as edges.

Finally, \citeauthor{pop00087}~\cite{pop00087} propose an ontology that describes the network resources and their connectivity. High-level policies are initially defined as statements, after which the developed tool leads its users trough policy refinement and generates low-level firewall rules. Policy refinement and rule generation are performed by applying automatic reasoning over the populated ontology.

\subsection{Categories outside the scope of this survey}
\label{sec-results-outside}

This subsection briefly describes initially collected papers that are outside the scope of this survey. Most such papers belong to the category \textit{rule analysis and error detection and correction}. The main goal of papers in this category is to analyze existing firewall policies or security policies specified on a higher level of abstraction, and make them easier to comprehend for firewall administrators and policy makers. Some of them can also highlight potential configuration errors and suggest corrections. One such approach is \citeauthor{pop00111}~\cite{pop00111}, where authors convert low-level firewall rules into a high-level formal representation that can be analyzed.

Many papers are concerned with testing, fuzzing, or formal verification of existing firewall policies. Formal verification approaches, such as \citeauthor{pop00489}~\cite{pop00489}, propose formal models that can be used to model firewall policies and automatically verify whether a firewall configuration satisfies its corresponding model. Testing and fuzzing approaches, such as \citeauthor{pop00482}~\cite{pop00482}, aim to generate test packets to check whether a given firewall configuration handles them as expected. The test packets are generated according to predefined criteria.

Several papers deal with problems regarding \textit{distributed firewalls and applications in cloud}.  One such approach is \citeauthor{pop00079}~\cite{pop00079}, where authors optimize the placement of firewall rules over multiple switches in a Software Defined Network (SDN).

In a number of papers, policies are defined on a higher level of abstraction than traditional firewall rules, but still retain a one-to-one correspondence to low-level technical details. As policies in this category are on a lower level of abstraction than papers in the scope of this survey, we label them as \textit{low-level rules}. In some other cases, such as \citeauthor{pop00060}~\cite{pop00060}, the policies are described on a sufficiently high level of abstraction, but are used directly on a custom firewall implementation rather than being compiled into low-level firewall rules. Such papers are labeled as \textit{firewall implementations} and are shown grouped together with low-level rules.

There are three categories representing other research areas. Papers labeled as \textit{Others - firewall related} present various subjects related to firewalls, those labeled as \textit{Others - non-English sources} are written in languages other than English, and the ones labeled as \textit{Others - not firewall related} deal with research unrelated to firewalls, such as CPU architectures.

\subsection{Comparable features of modern firewall systems}
\label{sec-results-industry}

\citeauthor{neupane2018next}~\cite{neupane2018next} recently published a focused survey of NGFWs and their capabilities. NGFWs combine extensive packet analysis with data from multiple sources, such as authentication logs, making them able to enforce policies aware of application-specific payloads and users' identities, and provide intrusion prevention capabilities \cite{pescatore2009defining}.
An extensive analysis of NGFWs is outside the scope of this paper. Instead, this subsection focuses on two features offered by Check Point Quantum NGFW \cite{checkpoint}, namely \textit{identity awareness} and \textit{application control}, which mirror common functionality proposed by the surveyed approaches. It must be noted that the authors of this sutvey are not in any way endorsed by Check Point and that this is only one of several NGFW systems frequently used in the industry.

Identity Awareness \cite{checkpoint} collects information about users and network resources from services such as \textit{Active Directory}, and associates network traffic with individual user accounts and devices. This enables straightforward implementation of policies that target individual users and devices instead of IP addresses. Such policies are much easier to maintain in modern organizations where users often work with multiple devices and connect trough wireless networks and virtual private networks (VPNs).

Application control \cite{checkpoint} involves deep packet inspection and is used to identify application specific network traffic. Consequently, admins can define policies that target individual applications regardless of ports, protocols, and other low-level technical details. Applications are detected using packet signatures from Check Point's internal database. Furthermore, applications are organized into categories and have an associated risk score \cite{checkpoint-db} that can be used when defining policies. For instance, a single firewall rule can filter all traffic related to applications inside the \textit{Anonymizer} category, such as VPN providers.

\section{DISCUSSION}
\label{sec-discussion}

The main goals of defining policies on a higher level of abstraction are to make them easier to comprehend and maintain, as well as to reduce the amount of domain knowledge required for the users who configure them. Most of the approaches relying on languages, graphical methods, and ontology certainly manage to improve policy comprehensibility. However, defining policies in approaches based on formal specifications requires extensive knowledge of formal logic and theorem proving, so these approaches effectively manage to replace the need for experienced network admins with a need for experts in formal methods, and consequently do not significantly improve comprehensibility.
% Formalne metode su loše jer su užasno nerazumljive xD

Important advantages of formal specification and ontology based approaches are that the high-level policies can be automatically checked for inconsistencies out-of-the-box and that the generated low-level firewall rules are guaranteed to adhere to the specification. Other types of approaches, such as \cite{pop00062}, can in some cases also include components that provide similar functionality, but we leave an overview of such functionality for future work.

% Grafički prikaz velikih mreža je jako težak i nepregledan
Another goal of high-level policy definition is to support transparent firewall management in large enterprise systems with numerous network segments and firewalls, which could require a very large set of rules to maintain. In this regard, most categories of approaches seem to, \textit{at least in theory}, support such application, with the exception of graphical approaches that are either highly specialized, like \cite{pop00013}, or are limited by the capabilities of their GUIs, like \cite{pop00059,pop00076}. The latter is caused by the inherent problem of visualizing large Enterprise networks, where visualizations can easily end up being very complicated and cluttered. A possible way of addressing this limitation would be to extend user interfaces of such approaches with abstraction of individual subsystems in the network.

In some approaches, information about users and resources can be imported from policy management software, such as Active Directory, while others require their users to define user identities and resources manually. A great advantage of the former is that resources and users can change over time, and this eliminates the need to handle such changes manually.

% Jezici i ontologije su problematični jer znanje treba održavati -> komercijalni sustavi to nude kao uslugu -> područje istraživanja je open-source katalog koji se kontinuirano održava
Most approaches suffer from a drawback that they need extensive configuration of domain specific data, such as descriptions of applications in \cite{pop00153}, to produce rules. As technologies evolve and resources are upgraded or replaced, knowledge bases and configurations must be continuously updated as well. As can be seen in Section \ref{sec-results-industry}, modern firewall systems already allow specifying rules that are aware of applications, users, and resources. Their vendors solve the aforementioned problem by delivering such firewalls together with additional services, often including continuous maintenance of the provided products and their knowledge bases.
%This type of delivery includes continuous maintenance of the provided products and their knowledge bases. 
Specifically, Check Point maintains and delivers a large collection of application signatures and offers a service trough which administrators can request assistance and report erroneous detection.
% Low level - prednost je da su dosta fleksibilne oko opisa tehničkih detalja, mana da treba puno tehničkog znanja i iskustva s firewallima - OVO SAM IZBACIO JER SE TEHNICKE DETALJE MOZE OPISATI U BAZI ZNANJA

% Komentar oko tvrdnji vs studija
% Situacija glede evaluacije - većinom nije pokazana praktična uporaba u industiji, manji broj studija
Last but not least, readers may observe that just six approaches performed validation of their results with domain experts, with the rest performing only case studies or performance evaluation. Proper validation with multiple domain experts is important as it can uncover practical problems with the approach and provide possible solutions.
%Firewall rule transcompilation... OUTSIDE THE SCOPE
% As noted by [CITATION], temporal rules are important for application in Industrial Control Systems. OUTSIDE THE SCOPE

\section{RELATED WORK}
\label{sec-related-work}

We found only one recent survey with a similar scope.
%\citeauthor{pop00238}~\cite{pop00238} are primarily concerned with detecting rule conflicts and do not compare high-level policies and methods of their translation.
\citeauthor{pop00049}~\cite{pop00049} comments some properties of approaches working with high-level policies, but does so very briefly, with the paper focusing primarily on detection of rule conflicts and anomalies. At the time of writing, \cite{pop00049} is over a decade old and misses the majority of approaches surveyed in this paper. In addition, this survey does not publish a methodology, making the reproducibility of its results challenging.
% In addition, none of the surveys above publish their methodology, making their reproduction challenging.

\section{CONCLUSION}
\label{sec-conclusion}

This paper provided a survey of approaches that aim to translate high-level definitions of security policies into low-level firewall rules. The surveyed approaches support definition of policies using either natural language, various domain languages, formal specification, or ontology. Each of them comes with its own advantages and disadvantages. Common drawbacks include the fact that the proposed high-level policy definition is often difficult to write and maintain, and that knowledge bases and configurations need constant maintenance. A potential solution to the latter would be to deliver such functionality trough services, in a similar manner to which vendors of NGFW systems like Check Point already deliver their products.

% There are several interesting directions for future work. 
% find more papers
% additional questions
% more thorough analysis

\section*{Acknowledgment}

% TODO: FINANCIRANJE NA PROJEKTU?
This work has been supported by the European Union's European Regional Development Fund, Operational Programme Competitiveness and Cohesion 2014-2020 for Croatia, through the project Center of competencies for cyber-security of control systems (CEKOM SUS), grant KK.01.2.2.03.0019.

\bibliographystyle{IEEEtranN}
\bibliography{references}

\end{document}